\documentclass[epj]{svjour}
\usepackage{graphicx}
\usepackage{graphics}
\begin{document}
\title{Influence of Chaos on the fusion enhancement by electron screening}
\author{S. Kimura\inst{1},  A. Bonasera\inst{1} and 
S. Cavallaro\inst{1}\inst{2}
}                     
%
%
\institute{
Laboratorio Nazionale del Sud, INFN, via Santa Sofia, 62, 95123 Catania, Italy\\
\and
Dipartimento di Fisica, Universit\`a degli Studi di Catania, via Santa Sofia, 64, 95123 Catania, Italy}
\date{Received: date / Revised version: date}
%
\abstract{
We study the effect of screening by bound electrons in low energy nuclear reactions.
We use molecular dynamics to simulate the reactions involving many electrons: 
D+$d$, D+D, $^3$He+$d$, $^3$He+D, $^6$Li+$d$, $^6$Li+D, $^7$Li+$p$, $^7$Li+H. 
Quantum effects corresponding to the Pauli and Heisenberg principles are 
enforced by constraints in terms of the phase space occupancy. 
In addition to the well known adiabatic and sudden limits,  
we propose a new "dissipative limit" which is expected 
to be important not only at high energies but in the extremely low energy region. 
The dissipative limit is associated with the chaotic behavior of the electronic motion.
It affects also the magnitude of the enhancement factor. 
We discuss also numerical experiments using polarized targets.   
The derived enhancement factors in our simulation are in agreement with those 
extracted within the $R$-matrix approach.    
%
} 
\maketitle
\section{Introduction}
\label{intro}
The relation between the tunneling process and dynamical chaos has been discussed 
with great interests in recent years.\cite{chaos,kb} 
Though the tunneling is completely quantum mechanical phenomenon, it is influenced 
by classical chaos.  In the sense that the the chaos causes the fluctuation of the classical 
action which essentially determines the tunneling probability. 
We study the phenomenon by examining the screening effect by bound electrons in the low energy 
fusion reaction. 
In the low energy region the experimental cross sections 
with gas targets show an increasing enhancement with
decreasing bombarding energy with respect to the values obtained by
extrapolating from the data at high energies~\cite{krauss}. 
Many studies attempted to attribute the enhancement of the reaction
rate to the screening effects by bound target electrons.
In this context one often estimates the 
screening potential as a constant decrease of the barrier height in the tunneling region through 
a fit to the data. A puzzle has been that the screening potential obtained by this procedure 
exceeds the value of the so called adiabatic limit, which is given by the 
difference of the binding energies of the united atoms and of the target atom
and it is theoretically thought to provide the maximum screening potential~\cite{rolfs95}.
Over these several years, the redetermination of the bare cross sections has been proposed 
theoretically~\cite{barker} and experimentally~\cite{junker}, using the Trojan Horse 
Method~\cite{thm,thm2,thm3}.
The comparison between newly obtained bare cross sections, i.e., astrophysical 
S-factors, and the cross sections by the direct measurements gives a variety of values 
for the screening potential. 
These values are often smaller than the sudden limit or larger than the adiabatic limit.
Theoretical studies performed 
using the time-dependent Hartree-Fock(TDHF) scheme~\cite{skls,ktab}  
suggest that the screening potential is between the sudden and the adiabatic limits.

One of the aims of this paper is to try to assess the effect of the screening quantitatively. 
Up to now, the dynamical effects of bound electrons have been studied only in some 
limited cases with a few bound electrons(the D+$d$ with atomic target~\cite{skls,ktab} and 
molecular D$_2$ target~\cite{smkls}, the $^3$He+$d$~\cite{skls}) with the TDHF method.  
We investigate here the dynamical effects, including the tunneling region, for other systems 
with many bound electrons; D+D, $^3$He+D, looking the effect of the electron capture of 
projectile. We see also some reactions including Li isotopes; $^6$Li+$d$, $^6$Li+D, 
$^7$Li+$p$ and  $^7$Li+H. 

To simulate the effects of many electrons, we use the constrained 
molecular dynamics (CoMD) model~\cite{kb,pmb,kb-2}.
At very low energies fluctuations are anticipated to play a substantial role. 
Such fluctuations are beyond the TDHF scheme.
Not only TDHF calculations are, by construction, cylindrically symmetric around the beam axis. 
Such a limitation is not necessarily true in nature and the mean field dynamics could be not 
correct especially in presence of large fluctuations.     
Molecular dynamics contains all possible correlations and fluctuations due to 
the initial conditions(events). For the purpose of treating quantum-mechanical systems 
like target atoms and molecules, we use classical equations of motion with
constraints to satisfy the Heisenberg uncertainty principle and the Pauli exclusion 
principle for each event~\cite{pmb}. 
In extending the study to the lower incident energies, we would like to stress the 
connection between the motion of bound electrons and chaos. 
In fact, depending on the dynamics, the behavior of 
the electron(s) is unstable and influences the relative motion
of the projectile and the target.  
The feature is caused by the nonintegrablility of the $N$-body system($N \ge 3$) and it is well known
that the tunneling probability can be modified by the existence of chaotic environment. 
We discuss the enhancement factor of the laboratory cross section in connection with 
the integrability of the system by looking the inter-nuclear and electronic oscillational 
motion.
More specifically we analyze the frequency shift of 
the target electron due to the projectile and the small oscillational motion induced 
by the electron to the relative motion between the target and the projectile. 
We show that the increase of chaoticity in the electron motion decreases the fusion probability.   

The paper is organized as follows.
In sect.~\ref{sec:form} we introduce the enhancement factor $f_e$ and 
describe the essence of the Constrained molecular dynamics approach briefly.
In sect.~\ref{sec:app} we apply it to asses the effect of the bound electrons during the 
nuclear reactions. We discuss also the relation between the amplitudes of the 
inter-nuclear oscillational motion and the enhancement factor.   
We summarize the paper in sect.~\ref{sec:sum}.

\section{Formalism}
\label{sec:form}

\subsection{Enhancement Factor}
We denote the reaction cross section at incident energy in the center of mass $E$ 
by $\sigma(E)$ and the cross section obtained in absence of electrons by $\sigma_0(E)$.
The enhancement factor $f_e$ is defined as 
\begin{equation}
  \label{eq:fenh}
  f_e\equiv\frac{\sigma(E)}{\sigma_0(E)}.
\end{equation}
If the effect of the electrons is well represented by the constant shift 
$U_e$ of the potential barrier, following~\cite{alr,skls}, 
($U_e \ll E$):
\begin{equation}
  \label{eq:fenh3}
  f_e\sim\exp\left[\pi\eta(E)\frac{U_e}{E}\right],
\end{equation}
where $\eta(E)$ is the Sommerfeld parameter~\cite{clayton}.

\subsection{Constrained Molecular Dynamics}
We estimate the enhancement factor $f_e$ numerically using molecular dynamics approach;
\begin{equation}
  \label{eq:rt}  
  \frac{d {\bf r}_i}{dt}= \frac{{\bf p}_i c^2}{{\mathcal E}_i}, \hspace*{0.5cm} 
  \frac{d {\bf p}_i}{dt}= -\nabla_{{\bf r}} U({\bf r}_i),
\end{equation}
where ({\bf r}$_i$,{\bf p}$_i$) are the position, momentum of the particle $i$
at time $t$.    
${\mathcal E}_i=\sqrt{{\bf p}_i^2c^2+m_i^2c^4}$, 
$U({\bf r}_i)$ and $m_i$ are its energy, Coulomb potential and mass, respectively. 
We set the starting point of the reaction at 10\AA~inter-nuclear separation. 
In Eqs.~(\ref{eq:rt}) we do not take into account the quantum effect of Pauli 
exclusion principle and Heisenberg principle.
As it is well known that these classical equations~(\ref{eq:rt}) can be derived by using the
variational calculus of Lagrangian ${\mathcal L}$ of the classical system as well.     
So as to take the feature of the Pauli blocking into account in this framework,  
we use the Lagrange multiplier method for constraints.

Our constraints which correspond to the Pauli blocking is 
$\bar f_i\le 1$ in terms of phase space density, note that the phase space 
density can be directly related to the distance of two particles, i.e.,
${\bf r}_{ij} {\bf p}_{ij}$, in the phase space.
Here ${\bf r}_{ij}=|{\bf r}_i-{\bf r}_j|$ and 
${\bf p}_{ij}=|{\bf p}_i-{\bf p}_j|$. 
The relation $\bar f_i \le 1$ is fulfilled, if ${\bf r}_{ij} {\bf p}_{ij}\ge
\xi_P\hbar\delta_{S_i,S_j}$, where $\xi_P=2\pi(3/4\pi)^{2/3}$. $i,j$ refer only to electrons and 
$S_i,S_j(=\pm 1/2)$ are their spin projection. 
For the Heisenberg principle ${\bf r}_{ij} {\bf p}_{ij}\ge\xi_H\hbar$, where $\xi_H=1$,
$i$ and $j$ refer to not only electrons but the nucleus.
It is determined to reproduce the correct energy of hydrogenic atoms. 
Obviously the conditions ${\bf r}_{ij} {\bf p}_{ij} =\xi_{H(P)}\hbar$ must be fulfilled 
in the ground state configuration rather than 
${\bf r}_{ij} {\bf p}_{ij} > \xi_{H(P)}\hbar$. 
 
Using these constraints, the Lagrangian of the system can be written down as 
\begin{eqnarray}
  {\mathcal L}=&&\sum_i \frac{{\bf p}^2_i c^2}{{\mathcal E}_i} - \sum_{i,j(\neq i)} U({\bf r}_{ij}) 
   + \sum_{i,j(\neq i)}\lambda^H_i \left( \frac{{\bf r}_{ij} {\bf p}_{ij}}{ \hbar}-1 \right) \nonumber \\
   +&& \sum_{i,j(\neq i)}\lambda^P_i \left( \frac{{\bf r}_{ij} {\bf p}_{ij}}{\xi_P \hbar}\delta_{S_i,S_j}-1 \right), 
\end{eqnarray}
where $\lambda^P_i$ and $\lambda^H_i$ are Lagrange multipliers.
The variational calculus leads 
\begin{eqnarray}
  \label{eq:rt2}  
  \frac{d{\bf r}_i}{dt}&=& \frac{{\bf p}_i c^2}{{\mathcal E}_i} 
  + \frac{1}{\hbar}\sum_{j(\neq i)}\left(\frac{\lambda_i^H}{\xi_H}
    +\frac{\lambda_i^P}{\xi_P}\delta_{S_i,S_j}\right) 
  {\bf r}_{ij}\frac{\partial {\bf p}_{ij}}{\partial {\bf p}_i }, \\
  \label{eq:pt2} 
  \frac{d {\bf p}_i}{dt}&=& -\nabla_{{\bf r}} U({\bf r}_i)
  - \frac{1}{\hbar}\sum_{j(\neq i)}\left(\frac{\lambda_i^H}{\xi_H}
    +\frac{\lambda_i^P}{\xi_P}\delta_{S_i,S_j}\right) 
  {\bf p}_{ij}\frac{\partial {\bf r}_{ij}}{\partial {\bf r}_i }. 
\end{eqnarray}
In order to obtain the atomic ground-state configuration, 
We perform the time integration of the eqs.~(\ref{eq:rt2}) and (\ref{eq:pt2}).
The value of $\lambda_i^H$ and $\lambda_i^P$ are determined depending on the 
magnitude of ${\bf r}_{ij}{\bf p}_{ij}$. If 
${\bf r}_{ij}{\bf p}_{ij}$ is (smaller)larger than $\xi_{H(P)} \hbar$, 
$\lambda$ has positive(negative) sign. Thus we change the phase space 
occupancy of the system.  
The constraints restrict us to variations 
$\Delta {\mathcal L}=0$ that keep the constraints always true~\cite{kb-2}.   
In this way we obtain many initial conditions which occupy different points in the phase 
space microscopically. 

In order to treat the tunneling process, we define the collective coordinates ${\bf R}^{coll}$ and 
the collective momentum ${\bf P}^{coll}$ as
\begin{equation}
  {\bf R}^{coll} \equiv {\bf r}_P-{\bf r}_T;   \hspace*{0.5cm}
  {\bf P}^{coll} \equiv {\bf p}_P-{\bf p}_T, 
\end{equation}
where ${\bf r}_T, {\bf r}_P$ ($ {\bf p}_T, {\bf p}_P$) are the coordinates(momenta) of the target 
and the projectile nuclei, respectively. 
When the collective momentum becomes zero, we switch on 
the collective force, which is determined by ${\bf F}_P^{coll} \equiv \dot{\bf P}^{coll}$ and 
${\bf F}_T^{coll} \equiv -\dot{\bf P}^{coll}$, to enter into imaginary time~\cite{bk}.
We follow the time evolution in the tunneling region using the equations,
\begin{equation}
  \label{eq:rti}  
  \frac{d {\bf r}_{T(P)}^{\Im}}{d\tau}= \frac{{\bf p}_{T(P)}^{\Im}}{{\mathcal E}_{T(P)}}; \hspace*{0.5cm} 
  \frac{d {\bf p}_{T(P)}^{\Im}}{d\tau}= -\nabla_{{\bf r}} U({\bf r}_{T(P)}^{\Im}) -2{\bf F}^{coll}_{T(P)},
\end{equation}
where $\tau$ is used for imaginary time to be distinguished from real time $t$.   
${\bf r}^{\Im}_{T(P)}$ and ${\bf p}^{\Im}_{T(P)}$ are position and momentum of the target
(the projectile) during the tunneling process respectively.   
Adding the collective force corresponds to inverting the potential barrier
which becomes attractive in the imaginary times.    
The penetrability of the barrier is given by~\cite{bk} 
\begin{equation}
  \label{eq:penet}
  \Pi(E)=\left(1+\exp\left(2{\mathcal A}(E)/\hbar\right)\right)^{-1},
\end{equation}
where the action integral ${\mathcal A}(E)$ is 
\begin{equation}
  {\mathcal A}(E)=\int_{r_b}^{r_a}{\bf P}^{coll}~d{\bf R}^{coll},  
\end{equation}
$r_a$ and $r_b$ are the classical turning points. The internal classical turning point 
$r_b$ is determined using the sum of the radii of the target and projectile nuclei.
Similarly from the simulation without electron, we obtain the penetrability of the bare 
Coulomb barrier $\Pi_0(E)$.

Since nuclear reaction occurs with small impact parameters on the atomic scale,
we consider only head on collisions. 
The enhancement factor is thus given by eq.~(\ref{eq:fenh}), 
\begin{equation}
  f_e=\Pi(E)/\Pi_0(E)
\end{equation}
for each event in our simulation. 
Thus we have an ensemble of $f_e$ values at each incident energy.

\section{Application to the Electron Screening Problem}
\label{sec:app}
\subsubsection{D+$d$ and D+D reactions}
\begin{figure}[htbp]
  \centering
  \includegraphics[width=8.3cm,clip]{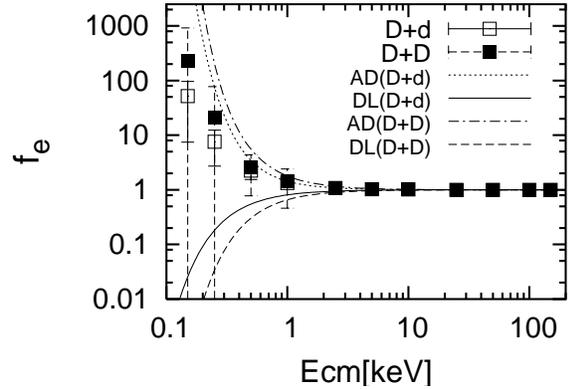}
  \caption{Enhancement factor as a function of incident center-of-mass energy 
    for the reactions D+$d$ and D+D. Error-bars represent the variances obtained from the events generated for each beam energy.}
  \label{fig:ef}
\end{figure}

Fig.~\ref{fig:ef}
shows the incident energy dependence of the enhancement factor for the reactions 
D+$d$ and D+D, where the systems involve 1 and 2 electrons respectively.
The open and closed squares show the average enhancement factors $\bar{f_e}$ 
over events for the reactions D+$d$ and D+D, respectively.   
The variances $\Sigma=\sqrt{\bar{f_e^2}- (\bar{f_e})^2}$ 
are shown with error bars. 
The dotted and dash-dotted curves show the enhancement factors
in the adiabatic limit $f_e^{(AD)}$ for an atomic deuterium target
and it is obtained by assuming equally weighted linear combination of 
the lowest-energy gerade and ungerade wave function for the electron, 
reflecting the symmetry in the D+$d$, i.e., 
\begin{equation}
  \label{eq:fe1}
  f_e^{(AD)}=\frac{1}{2}\left(
    e^{\pi\eta(E)\frac{U_e^{(g)}}{E}}+e^{\pi\eta(E)\frac{U_e^{(u)}}{E}}\right),
\end{equation}
where $U_e^{(g)}=$ 40.7 eV and $U_e^{(u)}=$ 0.0 eV~\cite{ktab,skls} for D+$d$ case. 
If we take into account the electron capture of the projectile, i.e.,
in the case of D+D, the enhancement factor in the adiabatic limit is 
\begin{equation}
  \label{eq:fe2}
  f_e^{(AD)}= \frac{1}{4}
  e^{\pi\eta(E)\frac{U_e^{(g.s.)}}{E}}+\frac{3}{4}e^{\pi\eta(E)\frac{U_e^{(1es)}}{E}},
\end{equation}
where $U_e^{(g.s.)}=$ 51.7 eV and $U_e^{(1es)}=$ 31.9 eV~\cite{kt}. 
The solid curve and dashed curve show the enhancement factors
in the dissipative limit $f_e^{(DL)}$ for the reactions D+$d$ and D+D respectively.
Notice how the calculated enhancement factor with their variances nicely 
ends up between the adiabatic and the dissipative limits.
We performed also a fit of our data using eq.~(\ref{eq:fenh3}) including 
the very low energy region and obtained $U_e=$ 15.9 $\pm$ 2.0 eV for D+$d$ case and 
$U_e=$21.6 $\pm$ 0.3 eV for D+D. 

Now we look at the oscillational motions of 
the particle's coordinates as the projection on the $z$-axis (the reaction axis). 
We denote the $z$-component of ${\bf r}_T, {\bf r}_P$ and ${\bf r}_e$ as 
$z_T, z_P$ and $z_e$, respectively.
Practically, we examine the oscillational motion of the electron around the target
$z_{Te}=z_e-z_T$ and 
the oscillational motion of the inter-nuclear motion, i.e., the motion between 
the target and the projectile, $z_{s}=z_T+z_P$, which 
essentially would be zero due to the symmetry of the system in the absence of the perturbation. 
\begin{figure*}
  \centering
  \includegraphics[width=11cm,clip]{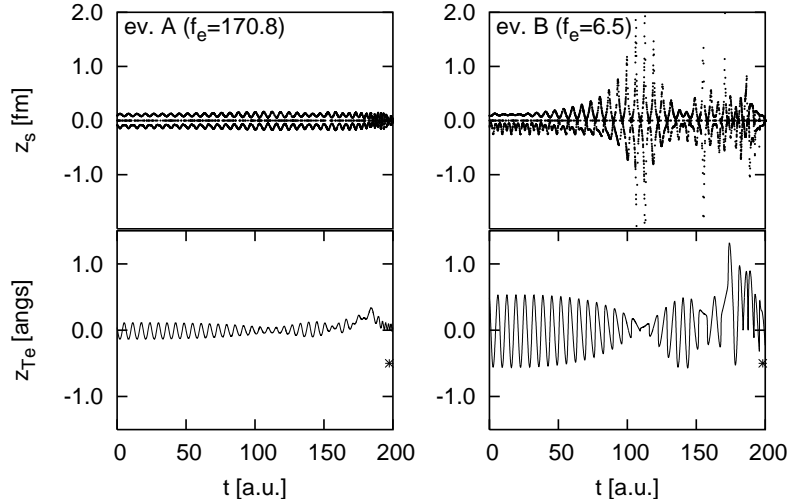}
  \caption{The oscillational motion of the electron around the target (lower panels)  
    and the inter-nuclear motion (upper panels) 
    as a function of time, in atomic unit, for two events, with large $f_e$(ev. A) 
    and small $f_e$(ev. B), for the D+d reaction at the incident 
    energy 0.15keV. The inter-nuclear separation is 10\AA~at $t=0$.}
  \label{fig:Rt}
\end{figure*}
In Fig.~\ref{fig:Rt} these two values are shown for 2 events, which have the enhancement 
factor $f_e=$ 170.8 (ev. A), and $f_e=$ 6.5 (ev. B), at the incident energy 
$E_{cm}=0.15$ keV.  
The panels show the $z_s, z_{Te}$ 
as a function of time. The stars indicate the time at which the system reaches the classical turning point.
It is clear that in the case of event B the orbit of the electron is much distorted from 
the unperturbed one than in event A.
Characteristics of $z_s$ are that (1) its 
value often becomes zero, as it is expected in the un-perturbed system, and 
(2) the component of the deviation 
from zero shows periodical behavior. It is remarkable that the amplitude of the deviation becomes 
quite large at some points in the case of event B which shows the small enhancement factor. 
Note that in event B one observes clear beats, i.e., resonances.   
Thus for two events, with the same macroscopic initial conditions, we have a completely different 
outcome, which is a definite proof of chaos in our 3-body system. 
We can understand these results in first 
approximation by considering the motion of the ions to be 
much slower than the rapidly oscillating motion of the electrons.~\cite{kb} 
From the Fig.\ref{fig:Rt} we can deduce the following 
important fact. If the motion of the electron is initially in the plane perpendicular to the reaction 
axis, the enhancement factor is large, event A(notice $|z_{Te}| \ll R_B$, i.e., the Bohr radius, at $t\sim 0$). 
On the other hand if there is a substantial projection 
of the electron motion, as in event B(the amplitude of $|z_{Te}|\sim R_B$ at $t\sim 0$), on the reaction 
axis the enhancement factor is relatively small
because of the increase of chaoticity. The fact suggests that if one performs experiments at very low 
bombarding energies with {\it polarized targets}, the enhancement factor can be controlled by 
changing 
the {\it polarization}. The largest enhancement would be gained with targets polarized 
perpendicularly to the beam axis.             

\begin{figure}
  \centering
  \includegraphics[width=8.3cm,clip]{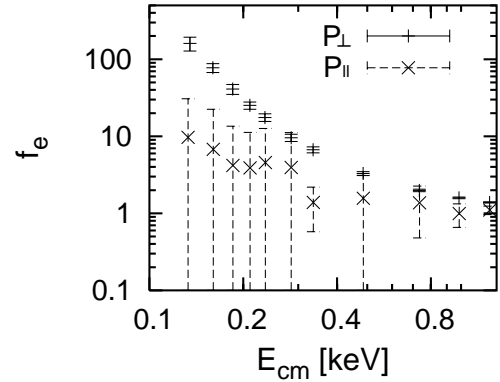}
  \caption{Incident energy dependence of the enhancement factor 
    for P$_{\perp}$ and P$_{\parallel}$ targets.}
  \label{fig:EFpol}
\end{figure}

In order to test this estimation, we prepared ensembles of target atoms which are polarized 
perpendicular(P$_{\perp}$) and parallel(P$_{\parallel}$) to the beam axis, numerically.   
In Fig.~\ref{fig:EFpol} we show the incident energy dependence of the average 
enhancement factor for the P$_{\perp}$ and P$_{\parallel}$ targets with 
pluses and crosses, particularly in the low energy region.  
The enhancement factors from the P$_{\perp}$ targets are always larger 
than that from the P$_{\parallel}$ targets. In contrast to the average enhancement 
from the P$_{\perp}$ targets, which increases monotonically as the incident 
energy becomes smaller, the average enhancement from the P$_{\parallel}$ targets 
fluctuates. It has also large variances at low energies. 
Remarkable thing is 
that with the parallel targets the enhancement factor often becomes less than 1. It means
that in this case the bound electron gives the effect of hindrance to the tunneling probability.

\subsubsection{$^3$He+$d$ and $^3$He+D reactions}

An excess of the screening potential was reported for the reactions $^3$He+$d$ with 
atomic gas $^3$He target, and D$_2$ + $^3$He with deuterium molecular 
gas target, for the first time in the reference~\cite{krauss}.   
Since then various experiments have been performed for these reactions. The incident 
energy covers from 5 keV to 50 keV for $^3$He+$d$. Though once the problem of the discrepancy between 
experimental data and theoretical prediction seemed to be solved by 
considering the correct energy loss data~\cite{lsbr}, recent measurements using measured 
energy loss data~\cite{aliotta} report larger screening potentials than in the adiabatic limit
for both reactions.      

The electron capture by the projectile plays a minor role in the case of $^3$He+d, since
electrons are more bound in helium targets. However 
in the recent measurement Aliotta et al. was performed using molecular D$_2^+$ and 
D$_3^+$ targets~\cite{aliotta}. Thus we assess the contribution from the 
reaction $^3$He+D, as well.     

\begin{figure}[htbp]
  \centering
  \includegraphics[width=8.3cm,clip]{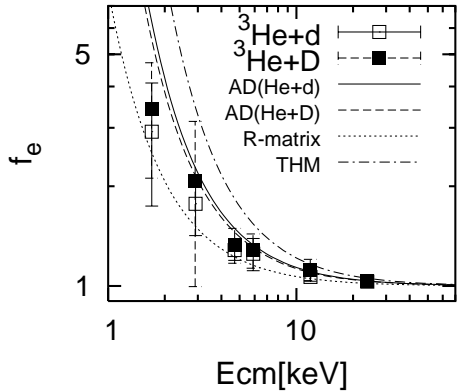}
  \caption{Enhancement factor as a function of incident center-of-mass energy 
    for the reactions $^3$He+$d$ and $^3$He+D.}
  \label{fig:ef3Hed}
\end{figure}

The enhancement factor in the adiabatic limit  give 
$U_e$=119 eV for $^3$He+$d$ and $U_e$=110 eV for $^3$He+D, respectively. 
These are shown in the figure~\ref{fig:ef3Hed} with the solid curve 
for $^3$He+$d$ and with the dashed curve for $^3$He+D.
The comparison of these two adiabatic limits implies that the electron capture 
of projectile would give a hindrance compared with the bare deuteron projectile.  
Meanwhile the latest analysis of the experimental data using $R$-matrix two level 
fit~\cite{barker} suggests the screening potential $U_e=$ 60 eV(corresponding 
enhancement factor is shown with dotted curve). 
The comparison between direct measurement and an indirect method, the Trojan Horse 
method, suggests the screening potential $U_e=$ 180$\pm$40 eV ( the corresponding 
enhancement factor is shown with dot-dashed curve)~\cite{thm3}.   
The average enhancement factors $\bar{f_e}$ over events in our simulations using the CoMD 
are shown with  
the open and closed squares for the reactions $^3$He+$d$ and $^3$He+D, respectively.   
The enhancement factors of the both reactions $^3$He+$d$ and $^3$He+D are in agreement 
with the extracted values using the $R$-matrix approach within the variances over 
all the events.  
Notice that our calculated enhancement factors for the two systems display an opposite trend 
as compared to the adiabatic limits.
The average enhancement factor of the reaction $^3$He+D agrees with the estimation of the 
adiabatic limit and the reaction $^3$He+d is below the corresponding adiabatic limit.  
The paradoxical feature comes from the fact that an electron between the two ions is often 
kicked out during the reaction process, i.e., the electron configuration seldom settles down 
the $^5$Li$^+$ ground state in the reaction $^3$He+d. It is known as autoionization in the 
context of the Classical 
Trajectory Monte Carlo method~\cite{gr}. Instead in the case of the $^3$He+D, the  deuterium 
projectile brings its bound electron in a tight bound state around the unified nuclei of 
$^3$He and $d$, practically it ends up with a ground state configuration of the $^5$Li atom.                 
The fits of the obtained enhancement factors suggests the screening potentials 
$U_e=$ 82.4 $\pm$ 1.9 eV for the $^3$He+$d$ and $U_e=$ 102.8 $\pm$ 3.0 eV for the $^3$He+D.

\subsubsection{$^6$Li+$d$, $^6$Li+D, $^7$Li+$p$ and $^7$Li+H}

The S-factors for the reactions $^6$Li+$d$, $^6$Li+$p$ and $^7$Li+$p$ were measured over 
the energy range 10 keV $< E_{cm} <$ 500 keV by Engstler,{\it et al.}~\cite{eknrsl}.
They used LiF solid targets and deuteron projectiles as well as deuterium molecular gas 
targets and Li projectiles. 

\begin{figure}[htbp]
  \centering
  \includegraphics[width=8.3cm,clip]{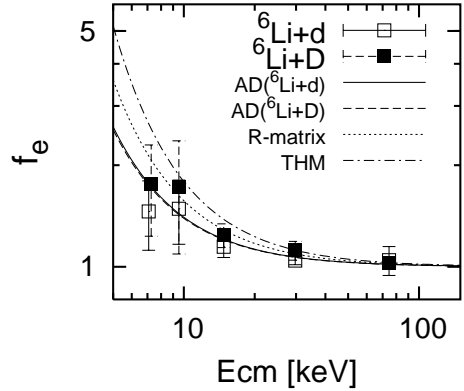}
  \caption{same as Fig.~\ref{fig:ef3Hed} but for the reactions $^6$Li+$d$ and $^6$Li+D.}
  \label{fig:ef6Lid}
\end{figure}

In the case of LiF target which is a large band gap insulator, one often approximates 
the electronic structure of the target $^6$Li($^7$Li) state by the $^6$Li$^+$($^7$Li$^+$) 
with only two innermost electrons. Thus for all three reactions one expects 
the screening potential in the adiabatic limit  $U_e^{(AD)}= 371.8-198.2\sim174$ eV. 
Instead if one uses the ground state of the  $^6$Li($^7$Li) atom and of the bare deuteron target 
as the initial state, $U_e^{(AD)}=$186 eV~\cite{bfmmq},  which is given by the solid curve  
in  Fig.~\ref{fig:ef6Lid} .

However one should be aware that the deuteron or hydrogen projectile plausibly moves with 
a bound electron in LiF solid insulator target~\cite{eder}.  
Under such an assumption we could estimate the screening potential 
$U_e^{(AD)}= 389.9-198.2\sim192$ eV.  In the case of molecular D$_2$ or H$_2$ gas targets, 
as well, we should consider  the electron capture by the   lithium projectile.  

The bare S-factors for the same 
reaction have been extracted using an indirect method, the Trojan-Horse Method through 
the reaction $^6$Li($^6$Li,$\alpha\alpha$)$^4$He~\cite{thm2}.
The comparison between direct and the indirect methods gives the screening potential 
$U_e=$ 320$\pm$50 eV. 
The corresponding enhancement factors are shown with the dash-dotted curve. 
The contrast between the direct measurement data and the theoretical estimation for 
the bare S-factor using the $R$-matrix theory gives $U_e$=240 eV. It is shown with dotted line.   
The extracted $U_e$ with the two different methods are larger than the 
adiabatic limit. 

\begin{figure}[htbp]
  \centering
  \includegraphics[width=8.3cm,clip]{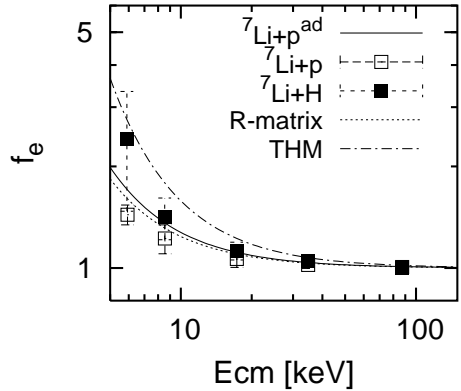}
  \caption{same as Fig.~\ref{fig:ef3Hed} but for the reactions $^7$Li+$p$ and $^7$Li+H.}
  \label{fig:ef7Lip}
\end{figure}

We simulate the reactions $^6$Li+$d$, $^6$Li+D, 
$^7$Li+$p$ and $^7$Li+H.
In the figure~\ref{fig:ef6Lid}(and \ref{fig:ef7Lip}) the open and closed squares show the enhancement 
factor for the reactions $^6$Li+$d$ and $^6$Li+D,(and $^7$Li+$p$ and $^7$Li+H) respectively. 

Again the average enhancement factors of the reaction $^6$Li+D($^7$Li+H) are larger than 
those of the $^6$Li+$d$($^7$Li+$p$). The enhancement factors of the reaction $^6$Li+D are in agreement 
with the extracted values using the $R$-matrix approach within the variances over all the 
events.  
The fit of the obtained average enhancement factors suggests the screening potentials 
$U_e=$ 152.0 $\pm$ 9.9 eV for $^6$Li+$d$ and $U_e=$ 214.4$\pm$18.5 for $^6$Li+D.
The screening potential for the reaction $^6$Li+$d$ in our simulation does not exceed the adiabatic 
limit nor extracted values using the $R$-matrix theory and THM, but one for $^6$Li+D verges on the
extracted values using the $R$-matrix approach.

\section{Summary}
\label{sec:sum}
We discussed the effect of the screening by the electrons in nuclear reactions 
at the astrophysical energies. 
We performed molecular dynamics simulations with constraints and imaginary time
for the reactions D+$d$, D+D, $^3$He+$d$, $^3$He+D, $^6$Li+$d$, $^6$Li+D, $^7$Li+$p$, $^7$Li+H. 
For all the reactions it is shown that both the average enhancement factors and their variances 
increase as the incident energy becomes lower. 
Using bare projectiles we obtained the average screening potential smaller than the 
value in the adiabatic limit for all reactions. It is because of the excitation or emission of 
several bound electrons during the reactions. 
The comparison between bare and atomic projectile cases for each reactions revealed that 
the electron capture of the projectile guides to larger enhancements.  
The derived enhancement factors in our simulation are in agreement with those 
extracted within the $R$-matrix approach including the variances over all the events.    

We report also the results of the numerical experiments using polarized targets for the reaction D+d.
Using P$_{\perp}$ targets we obtained relatively large enhancements with small variances, instead 
P$_{\parallel}$ target gives large variances of the enhancement factors and relatively small averaged 
enhancement factors. It is because with the P$_{\parallel}$ targets the force exerted from the electron 
to the relative motion of the nuclei is oscillational, in the direction of the beam axis, and the motion 
of the electron becomes often excited or unstable. It is the case where the chaoticity of the electron 
motion affects the tunneling probability and at the same time the enhancement factor of the cross section.     
This suggests that if one performs experiments at very low 
bombarding energies with polarized targets, the enhancement factor can be controlled by 
changing 
the polarization. The largest enhancement with targets polarized perpendicularly to the beam 
direction.             

%
%
%

\end{document}